\documentstyle[11pt,aaspp4]{article}
\begin{document}
\title{Detection of the Red Giant Branch Stars in M82 \\
Using the Hubble Space Telescope\footnote{Based on observations with the NASA/ESA 
{\it Hubble Space Telescope}, obtained at the Space Telescope Science Institute, 
operated by AURA, Inc. under NASA contract No. NAS5-26555.}}

\medskip
\author{\sc Shoko Sakai}
\affil{Kitt Peak National Observatory}
\affil{P.O. Box 26732, Tucson, AZ 85726 \hskip 0.5cm
shoko@noao.edu}

\author{\sc Barry F. Madore}
\affil{NED, California Institute of Technology, 
MS 100--22, Pasadena, CA 91125}
\affil{Observatories,  Carnegie  Institution of Washington,
813 Santa Barbara Street, Pasadena CA 91101 \hskip 0.5cm
barry@ipac.caltech.edu}

\bigskip
\centerline{Running Headline: {\it TRGB Distance to M82}}

\def\Deg{\hbox{${}^\circ$\llap{.}}}
\def\Min{\hbox{${}^{\prime}$\llap{.}}}
\def\Sec{\hbox{${}^{\prime\prime}$\llap{.}}}

\begin{abstract}

We present color--magnitude diagrams and luminosity functions
of stars in two halo regions of the irregular galaxy in M82, 
based on F555W and F814W photometry taken with the Hubble Space 
Telescope and Wide Field Planetary Camera 2.
The I--band luminosity function shows a sudden jump at I$\approx$23.95 mag,
which is identified as the tip of the red giant branch (TRGB).
Adopting the Lee et al. (1993) calibration of the TRGB based on the
RR Lyrae distances to Galactic globular clusters, we obtain the distance modulus
of $(m-M)_0 = 27.95 (\pm 0.14)_{\mbox{\tiny random}} [\pm 0.16]_{\mbox{\tiny
systematic}}$ mag. This corresponds to a linear distance of $3.9 (\pm 0.3)_{\mbox{\tiny random}}
[\pm 0.3]_{\mbox{\tiny systematic}}$ Mpc,
which agrees well with the distance of M81 deteremined from the HST observations
of the Cepheid variable stars.
In addition, we observe a significant number of stars apparently 
brighter than the TRGB.
However, with the current data, we cannot rule out whether these
stars are blends of fainter stars, or are indeed intermediate--age
asymptotic giant branch stars.

\end{abstract}

\it Subject headings: \rm galaxies: individual (M~82)   --
galaxies: irregular galaxies -- galaxies: distances
\bigskip

\begin{center}
{\it Accepted for publication in the Astrophysical Journal}
\end{center}

\newpage

\section{Introduction}

M82 has been one of the most frequently targeted 
candidates for studying sturburst galaxies and
is distinguished by its very high IR luminosity ($L_{IR} = 3 \times 10^{10}L_{\odot}$,
Telesco \& Harper 1980).
A cluster of young supernova remnants has also been observed around the nucleus
of M82 (Kronberg et al. 1985).
This galaxy is located in a small group of galaxies, which is comprised
of M81, M82, NGC~3077 and several other smaller dwarf galaxies.
The HI map of the M81 group revealed tidal tails bridging
M81 with M82 and NGC~3077, suggesting a recent interaction of these galaxies
(Yun, Ho \& Lo 1993).

The distance to the M81 group has been measured by Freedman et al. (1994)
using the $HST$ observations of Cepheid variables in M81.  
They report a distance modulus of $(m-M)_0 = $27.80 $\pm$ 0.20 mag.
Two other galaxies in the M82 group have also recently been the $HST$
targets.  Caldwell et al. (1998) observed dwarf ellipticals, F8D1 and BK5N,
and reported the distances measured by the tip of the red giant branch (TRGB)
method of 28.0 $\pm$ 0.10 and 27.9 $\pm$ 0.15 mag respectively.

As part of a long--term project to obtain direct distances to galaxies
in the nearby Universe using the TRGB method, 
we observed two fields in the halo of M82 using the HST and Wide Field Planetary
Camera 2.  
The details of observations and data reductions are reported in the following
Section 2.
In Sections 3 and 4, we discuss the detection of the RGB stars, and
report a distance using the I--band luminosity function.
In addition to the RGB stars, we detected a large number of 
stars brighter than the TRGB in the M82 halo regions.
We briefly explore in Section 5 what this population of stars may be.

\section{Observations and Reductions}

Two positions in the halo region of M82 were chosen 
for our HST observations.  A digital sky survey image of M82,
is shown in Figure~1, on which the 
$HST$ Wide Field Planetary Camera 2 (WFPC2) footprints are 
superimposed, indicating the two regions observed.
We refer to the region closer to the center of the galaxy as
Field I, and the other to the east as Field II.
The Planetary Camera (PC) chip covers the smallest area;
we refer to this as chip 1.
The three Wide Field (WF) chips cover the three larger fields and
are referred to as chips 2, 3 and 4 respectively,
counterclockwise from the PC.
A closeup $HST$ image of one of the chips, WF2 field of Field I, 
is shown in Figure~2.

Observations of the M82 halo region
were made with the WFPC2 on board the {\it Hubble Space Telescope}
on July 9, 1997 using two filters, F555W and F814W.
Two exposures of 500 seconds each were taken for both filters
at each position.
Cosmic rays on each image were cleaned before being combined to make a set
of F555W and F814W frames.  

The subsequent photometric analysis was done
using point spread function fitting packages DAOPHOT and ALLSTAR.
These programs use automatic star finding algorithms and then measure
stellar magnitudes by fitting a point spread function (PSF), constructed
from other uncrowded HST images (Stetson 1994).
We checked for a possible variation in the luminosity function as a function
of the position on each chip by examining the luminosity functions for
different parts of the chip.
For each frame, we find the identical luminosity function, 
confirming that there is no significant systematic offsets originating
from the adopted PSFs.

The F555W and F814W instrumental magnitudes were converted to the calibrated
Landolt (1992) system as follows.  (A detailed discussion is found in
Hill et al. 1998).  The instrumental magnitudes were first transformed
to the Holtzman et al. (1995) 0\Sec5 aperture magnitudes by determining
the aperture correction that need to be applied to the PSF magnitudes.
This was done by selecting 20--30 brighter, isolated stars on each frame.
Then all the stars were subtracted from the original image except for
these selected stars.  The aperture photometry was carried out for these
bright stars, at 12 different radii ranging from 0\Sec15 to 0\Sec5.
The 0\Sec5 aperture magnitudes were determined by applying the growth
curve analysis provided by DAOGROW (Stetson 1990), which were then compared
with the corresponding PSF magntiudes to estimate the aperture corrections
for each chip and filter combination.
The values of 
aperture corrections for each chip are listed in Table 1.
We use a different set of aperture corrections for two Fields.
Most of the values agree with each other within $2\sigma$,
however slight offsets between the corrections in the two Fields are
most likely due to the PSFs not sampling the images in the exactly same way.
When images are co--added, the combined images are not exactly identical
to the original uncombined images;
that is, the precise positions of stars on the frames are slightly
different. Thus we should expect
some differences in the aperture corrections of the same chip in 
two Fields.  

Finally, the 0\Sec5 aperture magnitudes are converted to the standard
system via the equation:
\begin{equation}
M = m + 2.5 \log t + {\mbox{C1}} + {\mbox{C2}} \times (V-I) + {\mbox{C3}} \times
(V-I)^2 + {\mbox{a.c.}},
\end{equation}
where $t$ is the exposure time, C1, C2 and C3 are constants and a.c. is the
aperture correction.
C1 is comprised of several terms including (1) the long--exposure WFPC2 magnitude
zero points, (2) the DAOPHOT/ALLSTAR magntidue zero point, (3) a correction
for multiplying the original image by 4 before converting it to integers
(in order to save the disk space), (4) a gain ratio term due to the difference
between the gain settings used for M82 and for the Holtzman et al. data 
(7 and 14 respectively), (5) a correction for the pixel area map which
was normalized differently from that of Holtzman et al. (1995), and
(6) an offset between long and short exposure times in the HST zero point
calibration.
C2 and C3 are color terms and are the same for all four chips.
In Table~2, we summarize all three constants for each chip.

\section{Detection of the Red Giant Stars in M82}
The $V$ and $I$ photometric results are shown in the color--magnitude
diagrams in Figure~3.
In Table~3, astrometric and photometric data for a set of brighter, 
isolated reference stars  are presented.  The X and Y coordinates 
tabulated refer to those
on the image of rootname u3nk0201m for Field I, and u3nka201m for Field II.
We also show luminosity function histograms in Figure~4.
In both Fields I and II, WF~4 field samples the least crowded
halo region of M82.
Based on the observations of Cepheids in M81, the parent galaxy of
M82, we know that the distance modulus of M82 is approximately 
$\mu_0 = 27.8$ mag.  Then the tip of the red giant branch should therefore
be observed at I $\sim$ 23.7 mag.
In all CMDs presented here, we can 
visually detect the position of the TRGB at $I \simeq 23.7 - 23.9$ mag
relatively easily, which is also evident in the luminosity functions,
as a jump in number counts, especially in those of Field I.
If we are observing the TRGB at around $I \sim 23.8$ mag, then
a significant number of brighter
stars are present in the halo regions of M82,
which are observed above the tip of the RGB in the CMDs.
In addition, comparing two regions, more of these stars are found in Field II.
This will be discussed more in detail in Section 5.

\section{TRGB Distance to M82}

The TRGB marks the core helium flash of old, low--mass stars which
evolve up the red giant branch, but almost instantaneously 
change their physical characteristics upon ignition of helium.
This restructuring of the stellar interior 
appears as a sudden discontinuity in the luminosity function and
is observed at $M_I \simeq -4$ mag in the $I-$band ($\sim 8200$\AA).
The TRGB magnitude has been shown both observationally and theoretically
to be extremely stable; it varies only by $\sim$0.1 mag for ages
2 -- 15 Gyr, and for metallicities between $-2.2 <$ [Fe/H] $< -0.7$ dex,
(the range bounded by the Galactic globular clusters).
Here, we use the calibration presented by Lee et al. (1993) which is
based on the observations of four Galactic globular clusters by
Da Costa \& Armandroff (1990). The globular cluster distances 
had been determined using the RR Lyrae distance scale based on 
the theoretical horizontal branch model for $Y_{MS}=0.23$ of 
Lee, Demarque and Zinn (1990), and corresponds to $M_V (\mbox{\small RR Lyrae}) = 0.57$ mag
at [Fe/H] = $-1.5$.

The top panel of each plot in Figure~5 shows
an $I-$band luminosity function smoothed by a variable Gaussian whose dispersion
is the photometric error for each star detected.
We apply a Sobel edge--detection filter to all luminosity functions to
determine quantitatively and objectively the position of the TRGB following
$E(m) = \Phi(I + \sigma_m) - \Phi(I - \sigma_m)$, where $\Phi(m)$ is
the luminosity function at magnitude defined at $m$, and
$\sigma_m$ is the typical photometric error of stars of magnitude $m$.
For the details of the Sobel filter application, readers
are referred to the Appendix of Sakai, Madore \& Freedman (1996).
The results of the convolution are shown as in bottom panels of 
Figure~5.
The position of the TRGB is identified with the highest peak in the filter output
function.

The TRGB method works as a distance indicator best in practice when the $I-$band
luminosity function sample is restricted to those stars in the halo
region only.  This is mainly due to three reasons: (1) less crowding,
(2) less internal extinction and (3) less contamination by
AGB stars which tend to smear out the ``edge'' defining the TRGB in the
luminosity function.
In Field I, the tip position is detected clearly in the luminosity
function and filter output for the WF~4 region at $I_{\mbox{\tiny TRGB}} 
= 23.82 \pm 0.15$ mag.
The one--sigma error here is determined roughly by estimating the ``full width
half maximum'' of the peak profile defining the TRGB in the filter 
output function.
In the WF~3 region, the tip is also observable at $I_{\mbox{\tiny TRGB}}
= 23.72 \pm 0.10$ mag,
slightly brighter than the case of WF~4. The simulations have shown that
the position of the tip shifts to a brighter magnitude due to crowding
effects (Madore and Freedman 1995), and that is what we observe on WF3.

The stellar population in Field II is comprised of more of these brighter stars
(which could be AGB stars),
thus restricting the luminosity function to the halo region helps
especially in determining the TRGB position.
Here, we obtain $I_{\mbox{\tiny TRGB}} = 23.71 \pm 0.09$ mag and
$I_{\mbox{\tiny TRGB}} = 23.95 \pm 0.14$ mag for WF~3 and WF~4 field respectively.
The tip magnitude of WF~3 agrees extremely well with that of the same
chip in Field I.  However, the TRGB magnitude defined by the WF~4 sample
is fainter by $0.17$ mag compared to the halo region of Field I.
There are several reasons to believe that the TRGB defined by the
Field II halo region would more likely correspond to the true distance of M82.
First, if one examines the WFPC2 image of Field I closely,
the presence of wispy, filamentary structures is recognizable.
Such features are likely to increase the uncertainties furthermore
due to variable reddening.  
Another but more important reason for putting less weight on the
Field I WF4 data is that there are far fewer stars observed
in this region.  
Madore and Freedman (1995) showed using a simulation that the population
size does matter in systematically detecting the TRGB position accurately.
That is, if not enough stars are sampled in the first bin immediately
fainter than the TRGB magnitude, the distance can be overestimated.
We show here again how the population sampling size affects our 
distance estimates.
We used $V$ and $I$ photometric data of the halo of NGC~5253 (Sakai 1999)
which is comprised of 1457 stars that are brighter than $M_I \leq -3$,
and is considered here as a complete sample.
The TRGB magnitude for this galaxy is $I = 23.90$ mag.
$N$ stars are then randomly selected from this NGC~5253 database
100 times, for which the smoothed luminosity function is determined.
The edge--detection filter is then applied to the luminosity function
in a usual fashion to estimate the TRGB magnitude.
This exercise was repeated for the case comprised solely of the RGB
stars; that is, the stars brighter than the TRGB were excluded from 
the parent sample.
We show the results for $N=20,100$ and $1000$ in Figure 6,
where the number distribution of TRGB magnitudes is shown for each simulation.
And in Table 4, we list the average offset from the TRGB magnitude.
In both cases, for smaller two samples, the TRGB determination becomes very 
uncertain, as the RGB population becomes indistinguishable from the brighter 
intermediate--age AGB population.
Or in the case where the RGB population is undersampled (the second scenario
in which only the RGB stars were included in the sample), the stars around
the tip of the RGB are missed, yielding an overestimated distance to this galaxy.  
Another way to present this effect is to plot the TRGB magnitude as
a function of the difference between the 0.15--mag bins immediately 
brighter and fainter than the TRGB.  This is shown in Figure 7.
For the least complete sample ($N=300$), the difference in number counts 
in the consecutive bins around the TRGB is merely $\sim$20.
This figure suggests that at least a number count difference of $\sim$40
is needed to estimate the TRGB position accurately.

Using the photometric data of the WF4 of Field II, the TRGB is detected
at $I = 23.95 \pm 0.14$ mag.
The foreground extinction in the line of sight of M82 is
$A_B = 0.12$ mag (Burstein and Heiles 1982).
Using conversions of $A_V / E(V-I) = 2.45$ and $R_V = A_V/E(B-V) = 3.2$ 
(Dean, Warren \& Cousins (1978), Cardelli et al. (1989) and Stanek (1996)),
we obtain $A_I = 0.05$ mag.  
To calculate the true modulus to M82,
we use the TRGB calibration of Lee et al. (1993), according to which
the tip distance is determined via the relation $(m-M)_I = I_{\tiny TRGB} -
M_{bol} + BC_I$, where both the bolometric magnitude ($M_{bol}$) and
the bolometric correction ($BC_I$) are dependent on the color of
the TRGB stars.  They are defined by: $M_{bol} = -0.19[Fe/H] - 3.81$ and
$BC_I = 0.881 - 0.243(V-I)_{\tiny TRGB}$.  The metallicity is in turn expressed
as a function of the $V-I$ color: $[Fe/H] = -12.65 + 12.6(V-I)_{-3.5} - 3.3(V-I)_{-3.5}^2$,
where $(V-I)_{-3.5}$ is measured at the absolute $I$ magnitude of $-3.5$.
The colors of the red giant stars range from 
$(V-I)_0 = 1.5 - 2.2$ (see Figure~4), which gives
the TRGB magnitude of $M_I = -4.05 \pm 0.10$.
We thus derive the TRGB distance modulus of M82 to be
$(m-M)_0 = 27.95 (\pm0.14)_{\mbox{\tiny random}} [\pm0.16]_{\mbox{\tiny 
systematic}}$ mag.
This corresponds to a linear distance of $3.9 (\pm 0.3) [\pm 0.3]$ Mpc.
The sources of errors include (1) the random uncertainties in the tip position
(0.14 mag) and (2) the systematic uncertainties, mainly those due
to the TRGB calibration (0.15 mag) and the HST photometry zero point (0.05 mag).
Unfortunately, because the TRGB method is calibrated on the RR Lyrae
distance scale whose zero point itself is uncertain at a 0.15 mag level,
the TRGB zero point subsequently has an uncertainty of 0.15 mag.
Recently, Salaris \& Cassisi (1997: SC97) presented a theoretical calibration
of the TRGB magnitude that utilized the canonical evolutionary models
of stars for a combination of various masses and metallicities for
$Y=0.23$ (Salaris \& Cassisi 1996).  SC97 find that their theoretical
calibration gives to a zero point that is $\sim$0.15 mag brighter than
the empirical zero point given by Da Costa \& Armandroff (1990).
They attribute this systematic difference to the small sample of stars
observed in the Galactic globular clusters.  We did find in previous
section that under-sampling the RGB stars 
leads to a systematically fainter TRGB magnitude, which seem to 
be in agreement with CS97.  Clearly, the issues pertaining to the
TRGB calibration need to be reviewed in detail in the future.
In this paper, we adopt the TRGB systematic calibration uncertainty 
of 0.15 mag based on these studies.

\section{Stars Brighter than the TRGB: What are they?}

It was noted in Figure~4
that the Field II appears to have a considerable number 
of stars that are brighter than the TRGB.
There are two possible scenarios to explain what these stars are:
(1) blends of fainter stars due to crowding, or
(2) intermediate--age asymptotic giant branch (AGB) stars.
To explain how much effect the crowding has on stellar photometry, 
we turn our attention to Grillmair et al. (1996) who presented 
HST observation of M32 halo stars.  They concluded that the AGB stars
detected in the same halo region by Freedman (1989) were mostly due to
the crowding.  Upon convolving the HST data to simulate the 0\Sec6 image
obtained at CFHT, they successfully recovered these brighter ``AGB''
stars.  While HST's 0\Sec1 resolution at the distance of M32, 770kpc, corresponds
to 0.37 pc, 0\Sec6 resolution at the same distance corresponds to 2.2pc.
Our HST M82 data has a resolution of 1.7 pc (0\Sec1 at 3.2 Mpc), indicating
that those stars brighter than the first--ascent TRGB stars are, by analogy
with M32, likely blends of fainter stars.  

If instead we were to adopt the second scenario in which these brighter
stars are actually AGB stars, the first striking feature in the CMDs shown 
in Figure 3 is that Field II contains significantly more AGB stars in comparison
to Field I.
In particular, we focus on the WF3 chip of each field;
we restrict the sample to smaller regions of WF3 chips where
the surface brightness is roughly in the range of $21.0 \leq \mu_i \leq 21.5$.
This corresponds to the lower 3/4 of WF3 chip in Field I (Regions 3A$+$3B in
Figure~8), and upper 3/4 of WF3 chip in Field II (Regions 3A$+$3B).
The difference in the number of AGB population of two Fields is compared
in term of $N_{AGB}/N_{RGB}$, defined here as the ratio of numbers of stars 
in a 0.5--mag bin brighter than
the TRGB to those in a 0.5--mag bin fainter than the TRGB.
We chose the 0.5--mag bin here as it might be less affected by the
incompleteness of stars detected at magnitudes $\sim$1 mag fainter than
the TRGB.
In calculating the ratios, we also assume that 20\% of the fainter giants
below the TRGB are actually AGB stars.
The ratios of Fields I and II are, respectively, $N_{AGB}/N_{RGB} = 58/193 = 0.30
\pm 0.04$ and $164/484 = 0.64 \pm 0.04$.
Restricting the samples furthermore to avoid the more crowded regions,
by using those stars in the section 3A only,
we obtain $N_{AGB}/N_{RGB} = 58/193 = 0.30 \pm 0.04$ and $87/172 = 0.51 \pm 0.06$
for Fields I and II respectively.
These ratios seem to suggest that the difference between the two fields is
significant, at a level of $4-5\sigma$'s.   
Because these subregions were chosen to match the surface brightness as closely
as possible, the blending of stars due to crowding should not be a major
factor in systematically making Field~II much richer in the intermediate--age
AGB population compared to Field~I.

Although the present analysis cannot by any means rule out 
crowding as the dominant effect and there is strong evidence that these brighter
stars above the TRGB are blends of fainter stars, we conclude the paper
by mentioning a possible connection between the presence of these brighter
stars (if real) with the HI distribution around this galaxy.
Yun, Ho and Lo (1993) presented the VLA observations of M82 which revealed
tidal streamers extending $\geq 10$kpc from M82, characterized by two
main structures.
One of these streamers extend northward from the 
NE edge of the galaxy, which coincides with our Field~II position.
The integrated HI flux map of Yun et al. does not, however, reveal any
neutral hydrogen in the region around Field~I.
If M82 is a tidally--disrupted system that has undergone direct interaction
with M81 and NGC~3077, could this have affected the star--formation history
of M82, enhancing a more recent star formation in the northeastern edge
of the galaxy (Field II)?
Answering this question is obviously beyond the scope of this paper,
requiring much deeper, higher--resolution observations, such as with
the Advanced Camera.

This work was funded by NASA LTSA program, NAS7-1260, to SS.
BFM was supported in part by the NASA/IPAC Extragalactic Database.

\newpage

\bigskip
\bigskip

\begin{center}
\begin{tabular}{cccc}
\multicolumn{3}{c}{\bf Table 1:  Aperture Corrections} \\
\hline
\multicolumn{1}{c}{Chip}  & 
\multicolumn{1}{c}{F555W}          & 
\multicolumn{1}{c}{F814W} \\
\hline
\multicolumn{3}{c}{Field I}\\
 WF 2  &  $-0.048 \pm 0.047$ & $-0.082 \pm 0.029$ \\
 WF 3  &  $+0.094 \pm 0.040$ & $-0.127 \pm 0.027$ \\
 WF 4  &  $+0.091 \pm 0.053$ & $+0.013 \pm 0.032$ \\
 & & & \\
\multicolumn{3}{c}{Field II}\\
 WF 2  &  $-0.039 \pm 0.038$ & $-0.029 \pm 0.015$ \\
 WF 3  &  $+0.053 \pm 0.040$ & $+0.185 \pm 0.022$ \\
 WF 4  &  $+0.155 \pm 0.063$ & $+0.005 \pm 0.024$ \\
 & & & \\
\hline
\end{tabular}
\end{center}

\bigskip
\bigskip

\begin{center}
\begin{tabular}{cccc}
\multicolumn{4}{c}{\bf Table 2: Transformation Coefficients} \\
\cr\hline
\hline\cr
%& & & \\
\multicolumn{1}{c}{Chip}&
\multicolumn{1}{c}{C1}&
\multicolumn{1}{c}{C2}&
\multicolumn{1}{c}{C3}\\
%& & \multicolumn{1}{c}{(sec)} & \\
\hline\cr
%\multicolumn{1}{c}{ }&
\multicolumn{4}{c}{F555W}\\
%\multicolumn{1}{c}{}\\
%  & & & \\
2 & $-0.957$ & $-0.052$ & $0.027$ \\
3 & $-0.949$ & $-0.052$ & $0.027$ \\
4 & $-0.973$ & $-0.052$ & $0.027$ \\
  & & & \\
%\multicolumn{1}{c}{ }&
\multicolumn{4}{c}{F814W}\\
%\multicolumn{1}{c}{}\\
%  & & & \\
2 & $-1.822$ & $-0.063$ & $0.025$ \\
3 & $-1.841$ & $-0.063$ & $0.025$ \\
4 & $-1.870$ & $-0.063$ & $0.025$ \\
  & & & \\
\hline\cr
\end{tabular}
\end{center}

\begin{deluxetable}{ccccccccc}
\tablenum{3}
\tablecolumns{9}
\tablewidth{0pc}
\tablecaption{Photometry of Reference Stars in M82}
\tablehead{
\colhead{ID} &
\colhead{Field} &
\colhead{Chip} &
\colhead{X} &
\colhead{Y} &
\colhead{RA (J2000)} &
\colhead{DEC (J2000)} &
\colhead{V} &
\colhead{I} 
}
\startdata
01 & I & 2 & 153.3 &  281.5  & 9:56:04.72 & 69:43:05.5  & 25.70 $\pm$ 0.15  & 23.52 $\pm$ 0.11  \cr
02 & I & 2 & 278.3 &  210.8  & 9:56:07.16 & 69:43:11.9  & 25.55 $\pm$ 0.21  & 23.16 $\pm$ 0.13 \cr
03 & I & 2 & 561.0 &  231.0  & 9:56:12.54 & 69:43:08.7  & 25.35 $\pm$ 0.22  & 23.96 $\pm$ 0.17  \cr
04 & I & 3 & 190.8 &  605.8  & 9:55:52.16 & 69:43:18.3  & 25.38 $\pm$ 0.15  & 23.78 $\pm$ 0.13  \cr
05 & I & 3 & 107.0 &  475.9  & 9:55:54.72 & 69:43:26.0  & 25.59 $\pm$ 0.19  & 23.91 $\pm$ 0.12  \cr
06 & I & 3 & 112.4 &  349.8  & 9:55:57.11 & 69:43:24.9  & 26.08 $\pm$ 0.19  & 23.90 $\pm$ 0.13  \cr
07 & I & 4 & 700.6 &  680.5  & 9:55:51.20 & 69:44:37.9  & 26.10 $\pm$ 0.17  & 24.09 $\pm$ 0.12  \cr
08 & I & 4 & 226.3 &  261.9  & 9:55:59.75 & 69:43:53.7  & 25.90 $\pm$ 0.17  & 23.86 $\pm$ 0.10  \cr
09 & I & 4 & 217.2 &  206.8  & 9:55:59.86 & 69:43:48.1  & 26.03 $\pm$ 0.20  & 24.27 $\pm$ 0.12  \cr
10 & II & 2 & 235.3 &  132.9  & 9:56:30.01 & 69:43:22.3  & 23.24 $\pm$ 0.10  & 23.09 $\pm$ 0.09  \cr
11 & II & 2 & 516.7 &  232.5  & 9:56:35.29 & 69:43:11.2  & 25.25 $\pm$ 0.25  & 23.34 $\pm$ 0.14  \cr
12 & II & 2 & 692.5 &  465.2  & 9:56:38.43 & 69:42:47.3  & 24.87 $\pm$ 0.18  & 24.24 $\pm$ 0.16  \cr
13 & II & 3 & 275.3 &  439.4  & 9:56:18.85 & 69:43:11.7  & 24.41 $\pm$ 0.10  & 23.47 $\pm$ 0.10  \cr
14 & II & 3 & 216.2 &  379.6  & 9:56:20.04 & 69:43:17.3  & 23.90 $\pm$ 0.09  & 23.08 $\pm$ 0.08  \cr
15 & II & 3 & 264.0 &  296.6  & 9:56:21.58 & 69:43:12.1  & 25.75 $\pm$ 0.17  & 23.70 $\pm$ 0.14  \cr
16 & II & 4 & 721.8 &  587.0  & 9:56:14.29 & 69:44:31.3  & 25.70 $\pm$ 0.18  & 24.29 $\pm$ 0.14  \cr
17 & II & 4 & 611.2 &  482.7  & 9:56:16.26 & 69:44:20.4  & 26.02 $\pm$ 0.25  & 23.99 $\pm$ 0.16  \cr
18 & II & 4 & 231.1 &  417.7  & 9:56:23.44 & 69:44:11.7  & 25.38 $\pm$ 0.21  & 23.59 $\pm$ 0.11 
\enddata
\end{deluxetable}

\begin{deluxetable}{ccc}
\tablenum{4}
\tablecolumns{3}
\tablewidth{0pc}
\tablecaption{TRGB Systematic Errors Due to
Undersampling in the RGB Population}
\tablehead{
\colhead{$N$} &
\colhead{Error} &
\colhead{Error}}
\tablehead{
\colhead{} & 
\colhead{(mag)}  &
\colhead{(mag)}
}
\startdata
 N  & RGB $+$ AGB &  RGB Only  \cr   
20  & $-0.24$  & $+0.08$ \cr 
50  & $-0.24$  & $+0.04$ \cr 
100 & $-0.20$  & $+0.03$ \cr
200 & $-0.11$  & $+0.01$ \cr
500 & $-0.04$  & $+0.00$ \cr
1000 & $-0.01$ & $+0.00$   
\enddata
\end{deluxetable}

\newpage

{\bf Figure Captions}

Figure 1: A digital sky survey image of M82.
Two footprints indicate the regions of the $HST$ WFPC2 observations.

Figure 2: A closeup view of a WFPC2 field of Field~II, WF~2 chip.

Figure 3: $(V-I) - I$ color magnitude diagrams of each chip in two fields.
Note a significant number of stars brighter than the TRGB are observed in
WF~2 and WF~3 chips, especially in Field~II.
The arrow in each CMD shows the location of the TRGB, while
the dotted line represents $V = 26.9$, roughtly indicating the incompleteness level.

Figure 4: I--band luminosity histograms for each chip in both fields.

Figure 5: Smoothed luminosity function (top) and edge--detection filter
output for each position.

Figure 6: Number of simulations detecting the TRGB magnitude plotted on
the x--axis.  See text for details.

Figure 7: The observed TRGB magnitude in each simulation plotted as a function
of the difference between the number of stars in the 
1.5--mag bin above and below the TRGB ($N_+ - N_-$).

Figure 8:  A schematics showing the regions used to calculate the
ratios of AGB--to--RGB stars.

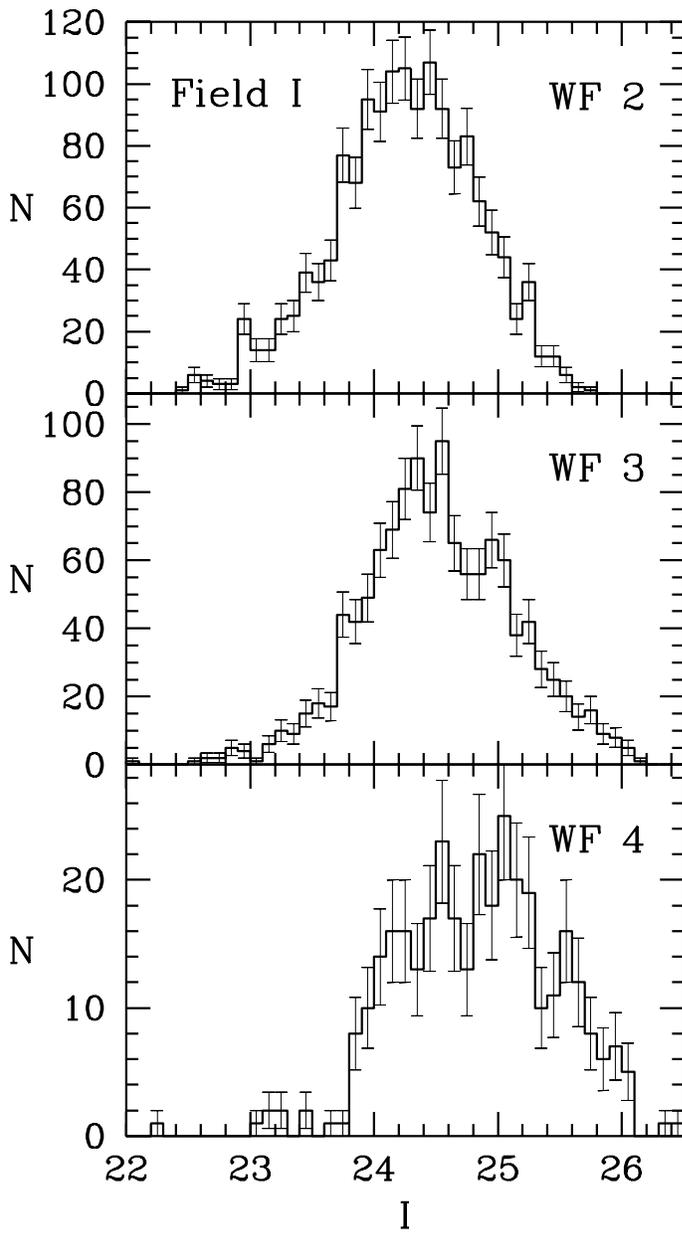
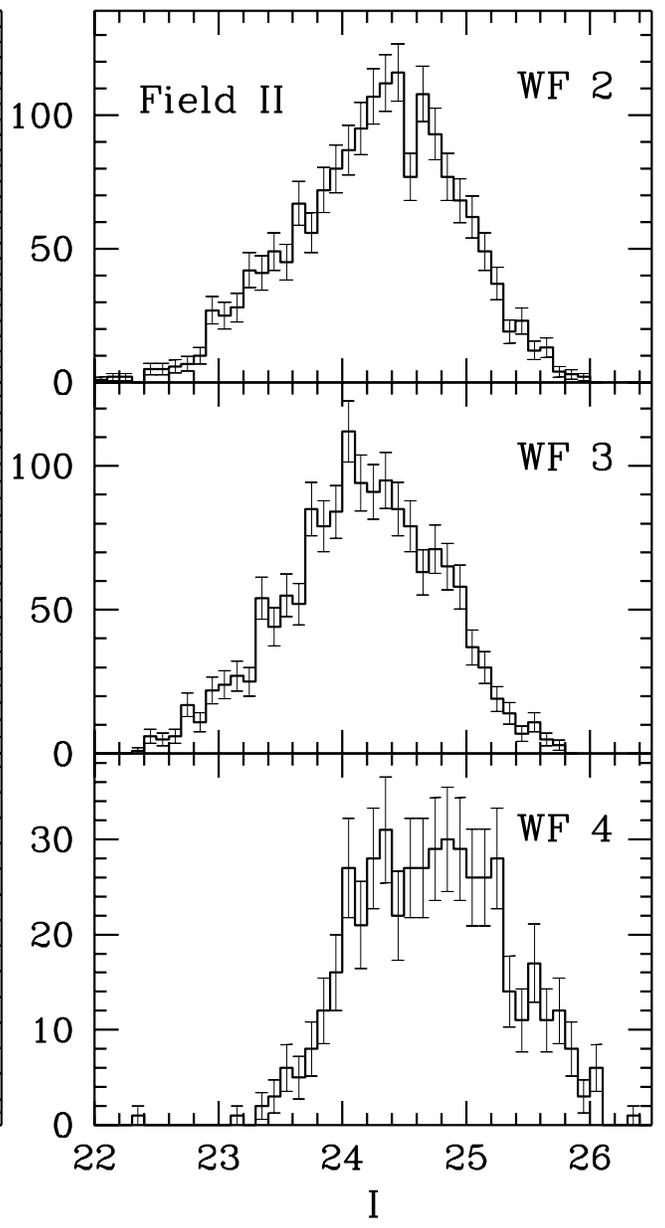
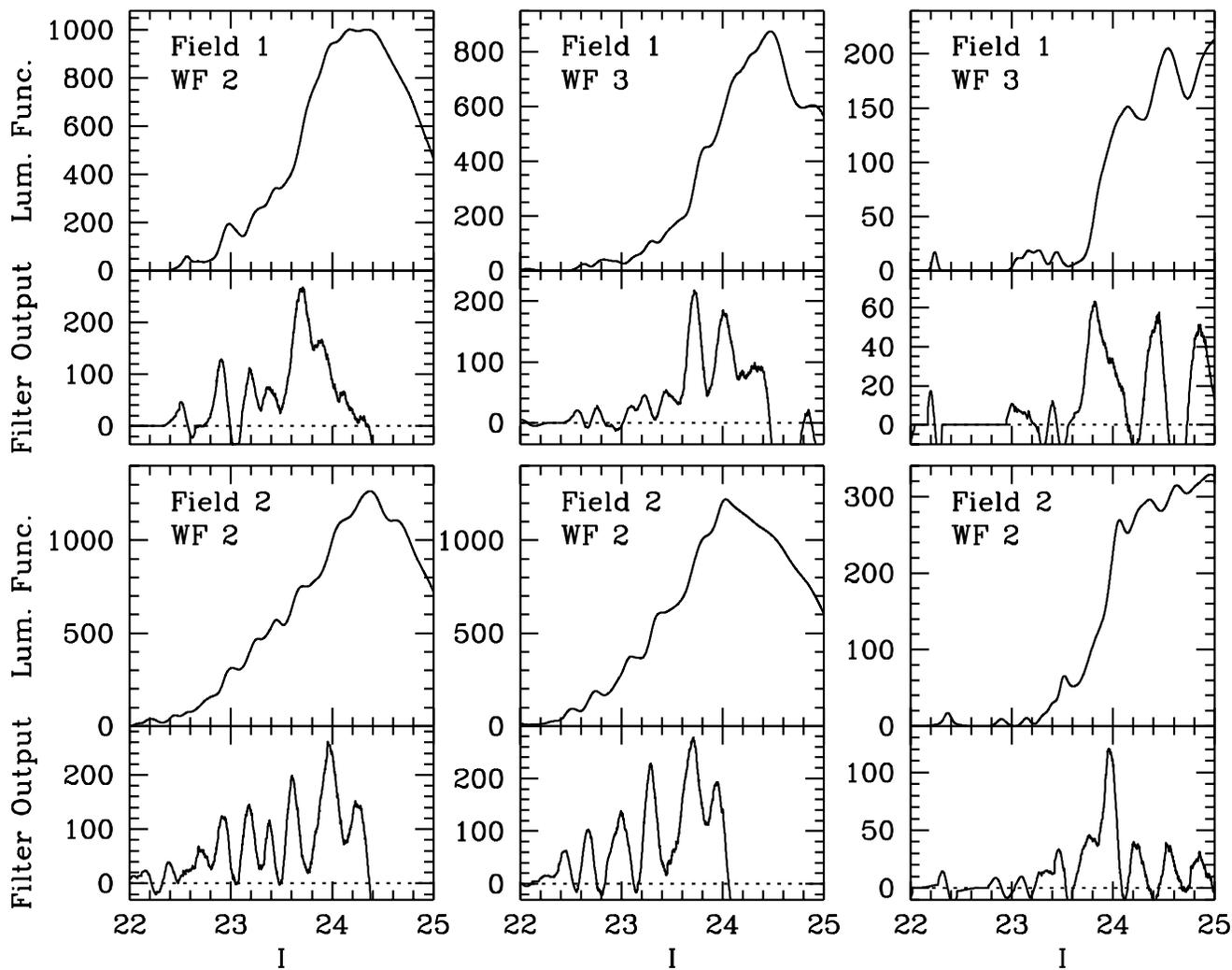
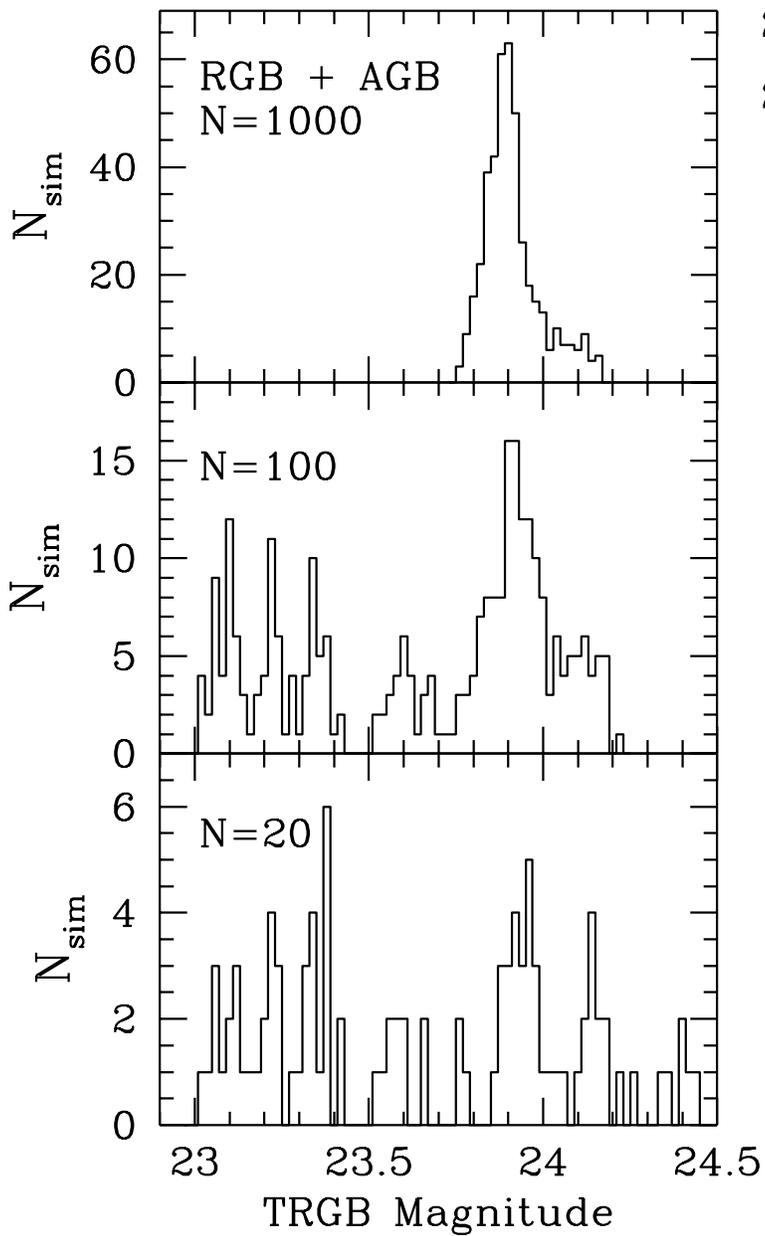
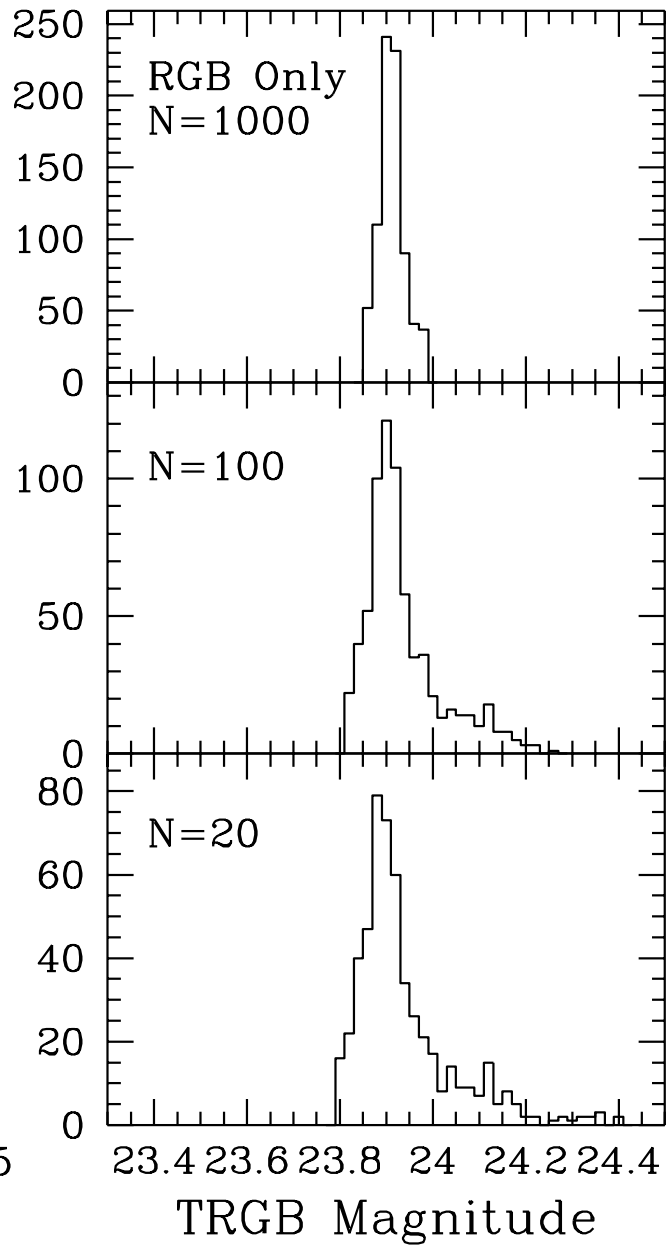
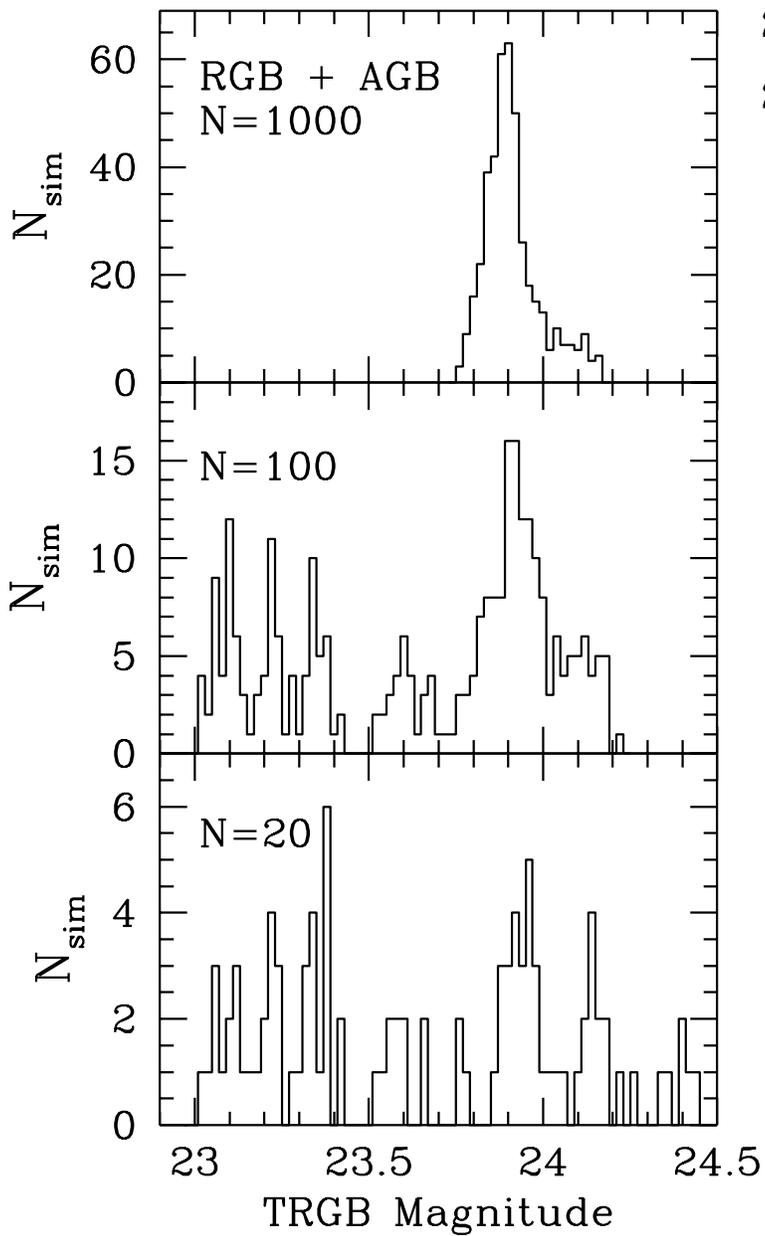
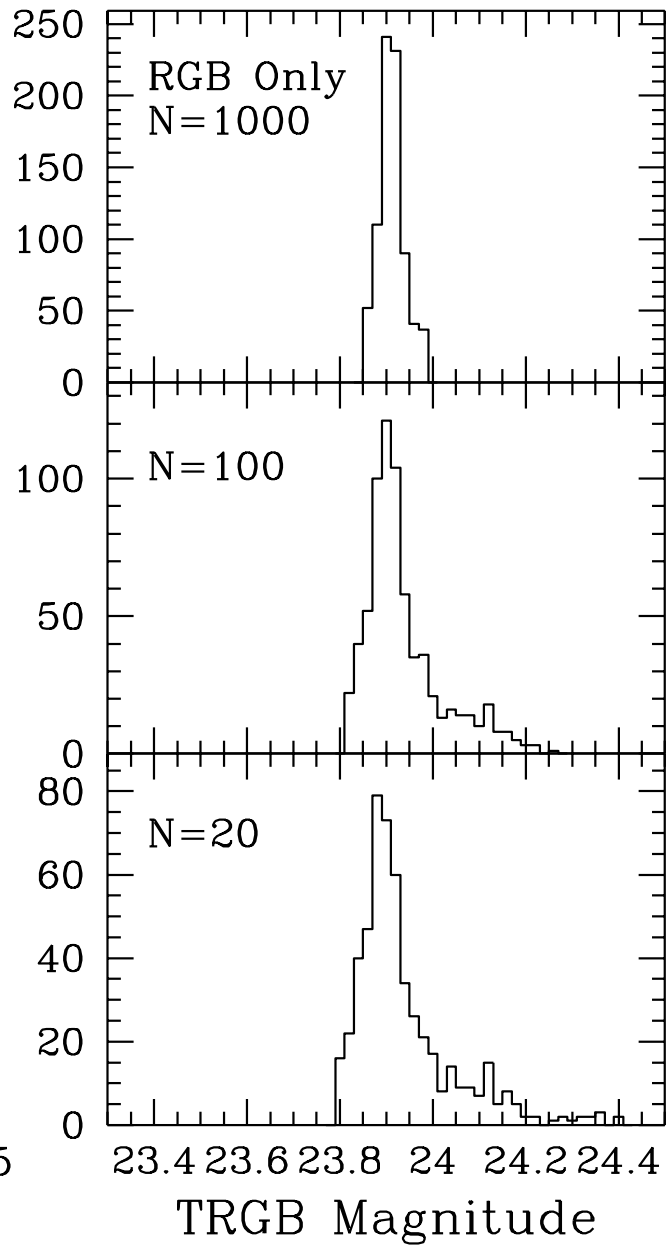
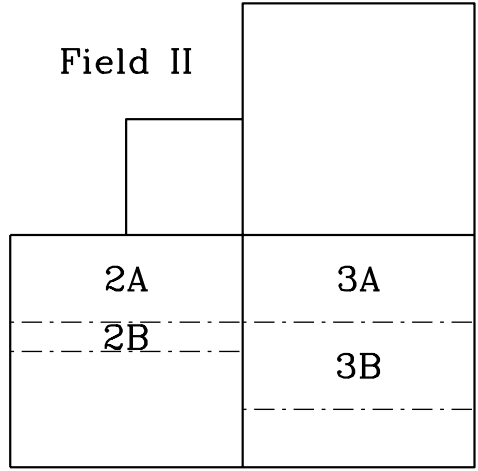
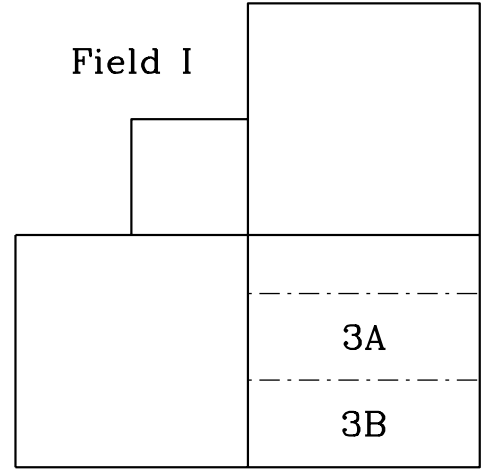
\end{document}